# Optimization Techniques to Improve Inference Performance of a Forward Propagating Neural Network on an FPGA.

Matthew J. Adiletta
Student, Worcester Polytechnic Institute
100 Institute Road,
Worcester, Massachusetts
mjadiletta@wpi.edu

Brian Flanagan
Student, Worcester Polytechnic Institute
100 Institute Road,
Worcester, Massachusetts
bflanagan@wpi.edu

*Abstract:* This paper describes an optimized implementation of a Forward Propagating Classification Neural Network which has been previously trained. The implementation described highlights a novel means of using Python scripts to generate a Verilog hardware implementation. The characteristics of this implementation include optimizations to scale input data, use selected addends instead of multiplication functions, hardware friendly activation functions and simplified output selection. Inference performance comparison of a 28x28 pixel "hand-written" recognition NN between a software implementation on an Intel i7 vs a Xilinx FPGA will be detailed.

*Keywords: Feed Forward Neural Network; FPGA Neural Network Implementation; Inference Performance Testing.*

## I. INTRODUCTION

Artificial Intelligence is a broad field. Intelligent agents have been developing in complexity since the mid 1900's. The development of the Neural Network (NN) in 1943 - accredited to Warren McCulloch and Walter Pitts - began one of the world's greatest research areas for intelligent agents. At the time of their development, computers could not meet the computational demands required by neural networks, limiting further advancements. In years following, the ability for computers to process large amounts of data and execute calculations increased exponentially. Today, computers are able to processes incredible amounts of data in reasonable amounts of time. The ability to process data quickly and efficiently is the main reason that artificial neural networks are emerging and ubiquitous today.

### A. Introduction to Biological and Articial Neural Networks

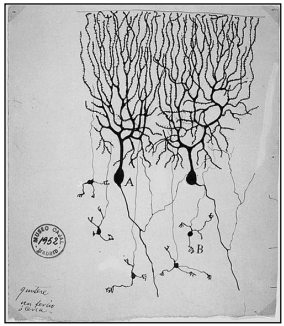

Figure 1: Drawing by Santiago Ramon y Cajal (1899) - interpretation of a system of neurons in the brain.

Neural networks are a computer system modeled after the human brain. Figure 1 below shows the fundamental principal of biological neurons.

At a high level, neurons in the brain take in signals then output signals. The input values to a neuron are called dendrites, the output value of a neuron is called an axon and the connections between neurons are called synapses. If the input signal to a dendrite is higher than a certain threshold, the neuron becomes active and will output a signal to other neurons. The output signal of a single neuron can be linked to thousands of other neuron dendrites. This network of connected signals provides for a very complicated system capable of intelligence.

It is the goal of computer scientists researching intelligence to represent a computer system in the same form as a biological brain, hoping to create an intelligent agent.

An artificial neural network is structured similarly to a biological brain. To compare the input and output structure of neural networks and neurons, a node in a neural network is the equivalent to a neuron. The connection between nodes is explicitly calculated using weight matrices, equivalent to a neuron's synapses. The input and output of a node are equivalent to the dendrites and axons.

One main difference between a classical artificial neural network and a biological system is the level of complexity. Classical artificial neural networks are structured in interconnected layers such as in figure 2 below. Figure two below shows a three layer, fully connected, neural network composed of no looping connections or overlapping nodes. Despite the simplicity of the architecture, rudimentary intelligence is possible.

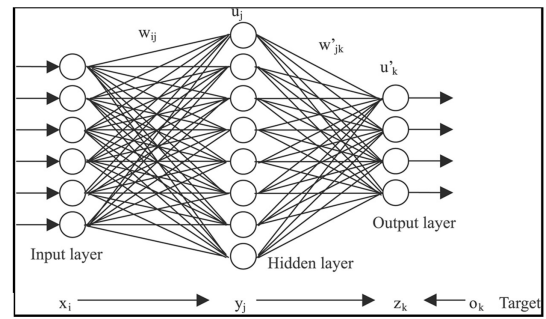

Figure 2: Classic three-layer feed forward Neural Network. Neurons are identified with circles. Connections are indicated with lines.

### B. Flow of a Neural Network

This section will describe how an input propagates through a neural network to produce an output. First an input must be provided to the neural network. In figure two, this is called the "input layer". For the input to reach the next "hidden layer" in figure 2, the input must pass through a fully interconnected, scaling and summing function. Each input node is directly connected to each node in the "hidden layer" by a scaling factor called a weight. This means that each input is scaled by the weight associated between the input node and the corresponding hidden node. Then all scaled input values to a single hidden node are summed to calculate a single input value to the hidden node.

The hidden node then uses an activation function to determine the output of the node. Various activation functions are explained in future sections of this paper. For simplicity, a basic comparator activation function will be used for this explanation. If the sum of all weighted input is higher than a predefined threshold value, then the resulting hidden node output value is high, otherwise, the output is low.

The process of feeding node output forward to the next layer as an input along with a given weight between nodes is repeated between the hidden layer and the output layer – the hidden outputs are scaled by a weight value and summed at each output layer node. The output layer produces an output value for each node. The final output represents the neural network "prediction."

*C. Training vs. Inference*

When architecting an informed neural network, two phases occur. The first phase is training, and the second phase is testing the quality of the resulting "inference." Training and testing/inference are completely independent of each other.

A neural network must be able to take an input and make an accurate prediction. To form an accurate prediction, the neural network must be trained. Training occurs by passing an input through the network and comparing the prediction with the actual values. The error is then propagated back through the neural network. Since the only constants in a neural network are the weight matrices, the error is only used to update the weight matrices. The result of updating the weights is hopefully an increase in prediction accuracy. The inputs used to train the neural network must be completely separate from the testing datasets.

To calculate the accuracy of a trained neural network, the neural network must be tested. To test a neural network, a new set of inputs must be passed through the network. Each predicted result is compared against its corresponding actual value. The percent correctly predicted is the accuracy of the neural network.

The datasets for training and testing must be different. The reason for this is the neural network must not have already "seen" the exact input data before. A comparable example is if a student is studying for a test and the professor shows the student all the questions and answers beforehand. The student is biased while taking the exam. Presumably, the final grade on the student's exam will be higher than on an exam where the student had not already seen the questions and answers. This same principle holds for neural networks.

To properly determine the accuracy of a neural network, a completely separate dataset must be used to test the neural network.

*D. Activation Functions*

The general structure of a neural network can make a big difference in the accuracy of the network. Two main design decision made that effect the accuracy of the network is the number of layers in the network and the number of nodes per layer. Another critical design choice of a neural network is the type of activation function. As previously stated, the activation function is used to translate an input into an output. In the previous example, a simple comparator was used with a threshold value.

There are a variety of functions that can be used as activation functions. These are a few examples: step function, biased step function, linear function, rectilinear function, sigmoid, and tanh.

The research presented in this paper uses the step function and sigmoid function as activation functions. Figure three below shows a comparison between the two functions.

These two functions are similar in nature because they both have an inflection point when the input is zero. Furthermore, large positive input values result in values close to one, and large negative input values all result in values close to zero.

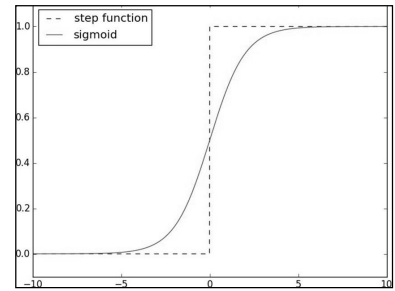

Figure 3: Sigmoid Function vs Step Function.

*E. What is a Classification Neural Network?*

A classification neural network is a type of network that groups inputs based on shared features and characteristics. The predicted group associated with an input is determined by the highest active output node of the neural network.

An example of a classification neural network is a number classifier. If an image of a number is provided to a number classifier agent, the agent will derive the group that the imaged is associated with. For a simple number classifier, the possible groups are integer numbers between zero and nine.

*F. Research for modern Neural Networks*

Today, the research on neural networks is extensive. Research is being conducted to improve accuracy, speed, and complexity of neural networks. The following research proposes a novel hardware architecture for a feed forward classification neural network implementation.

Modern neural networks can receive an input and derive an output in the time interval of milliseconds via software. In modern day however, milliseconds are still significant compute time intervals for a computer, especially given modern computers operate in gigahertz frequencies. Furthermore, neural networks are being used to solve real-time problems such as object detection for self-driving cars or responsive conversation recommendations where humans have limited patience for waiting.

Let's look at a situation of a self-driving car. If a software based neural network requires tens or hundreds of millisecond prediction latency, then the fastest reaction time of a car agent is limited by the speed of the neural network, plus the actuating mechanical constructs. Creating deeper neural networks with higher accuracy only increases this latency. Autonomous car system makers are addressing this latency in order to improve safety for self-driving cars by employing hardware based neural networks.

The following sections describe a novel neural network hardware architecture that more closely resembles a biological brain than classical neural networks for two reasons: (1) much faster execution time by maximizing parallel computation (2) asynchronously passing data between nodes.

## II. SOLUTION TO IMPROVE NEURAL NETWORK PREDICTION LATENCY

To reduce prediction latency within a neural network, all computations within layers should be executed simultaneously which is also known as task parallel execution. Task parallel is impossible on modern Von Neuman linear computation computers. A different architecture must be employed.

A programmable hardware construct which is capable of task parallel computation is a Field Programmable Gate Array (FPGA). FPGA's are suitable for this problem because they can be programmed to generate specific hardware to execute many operations at once.

Neural networks have already been implemented on FPGA's, however, modern implementations all have drawbacks. For instance, most implementations are using expensive hardware to create massive Look Up Tables (LUTs) that model complicated activation functions such as the sigmoid function. Furthermore, these implementations use a clock to synchronize events. The research presented in this paper implements an asynchronous hardware architecture for a classification neural network which avoids the use of deep LUTs.

### A. Example Implementation of a Three Layer Neural Net

To implement a neural network on an FPGA we first modeled the system in python. To begin, a simple three-layer neural network is created, sampled from the book "Make Your Own Neural Network" by Tariq Rashid.

The input to this neural network is a 28x28 image from the MNIST dataset. The result of this neural network is a classification for the input image as a number between zero and nine.

The neural network that is setup has 784 input nodes (28x28 image vectorized), 500 hidden nodes, and 10 output nodes. There are two weight matrices, 784x500 and 500x10. The activation function is a sigmoid function, with output values between zero and one.

The neural network is trained via standard back propagation techniques for updating weights. After training the neural network on 1000 images with 5 epochs, the weights are saved and stored so that the neural network can be loaded at any time and tested.

The accuracy of the trained neural network is 98% on testing data from the MNIST dataset. These results can be reproduced following the techniques described in Tariq Rashid's book, "Make Your Own Neural Network" stated above.

Our next step is to make adjustments to the neural network so that it can be implemented on an FPGA such that the architecture generated is asynchronously communicating and avoiding deep LUT use.

## III. WHAT HAPPENS WHEN COMPLEXITY IS REDUCED?

Field Programmable Gate Arrays are very simple in principle. They have logic and memory units and can be programmed to form combinational or sequential logic. FPGAs have a hard time with a few techniques that CPU's excel at such as decimal precision and floating-point arithmetic. FPGA's also have a hard time with error detection, debugging, timing constraints, and more – mostly problems that computer programmers take for granted.

That being said, the neural network realized in python is clearly leveraging use of floating-point arithmetic and vectorized calculations. The activation function itself only output values between zero and one. These are problems that must be addressed. Note that the changes listed below only effect inference testing as the training has already solidified the structure of the neural network.

### A. Simplifying the Activation Function

The first simplification of our feed forward classification neural network is to change the activation function from a sigmoid function to a step function. All other components of the neural network remain the same.

The resulting accuracy for the neural network after making this change remains surprisingly high at 95%. Only a 3% decrease in accuracy despite a very large decrease in activation precision. This optimization is great for the neural network implementation because of the simplicity of the step function – it is essentially a comparator. If the input value is larger than a threshold, the output is a one, and if the input value is less than a threshold, the output is a zero. This is easily implemented on an FPGA using a simple shallow look up table.

### B. Simplifying Input Values to Either 0 or 1

The next identified problem is with the input values. Recall, there is a 28x28 (784pixels) input image matrix of a "hand-written" number. Each pixel value is between 0 and 256. This is the input to the neural network which is then scaled between 0 and 1. This scaling is again a problem for an FPGA that cannot easily process floating point numbers between 0 and 1. The solution here is to compare the raw input value between 0 and 256 with a cutoff value of 128. If the input value is larger than 128, then the input is represented as a one, otherwise it is zero.

The resulting accuracy decreased from 95% to only 94% with a net decrease in accuracy of only 4% from the optimal neural network setup. This is another promising simplification that is easily implemented on an FPGA.

### C. Define All Weight Values as Integers

The final optimization needed to completely remove decimal precision from the neural network is to transform the weight matrices. These matrices are not acceptable with their current decimal precision. The solution is to cast all the weights to integers. Testing the neural network with this optimization resulted in a new accuracy of 92%, only a 6% decrease from the original neural network accuracy of 98%.

This is incredible, purely because it means that decimal precision on a neural network only adds about 6% accuracy. A six percent decrease in accuracy is a relatively small price to pay compared to the expense of complicated hardware and slowness of floating-point precision.

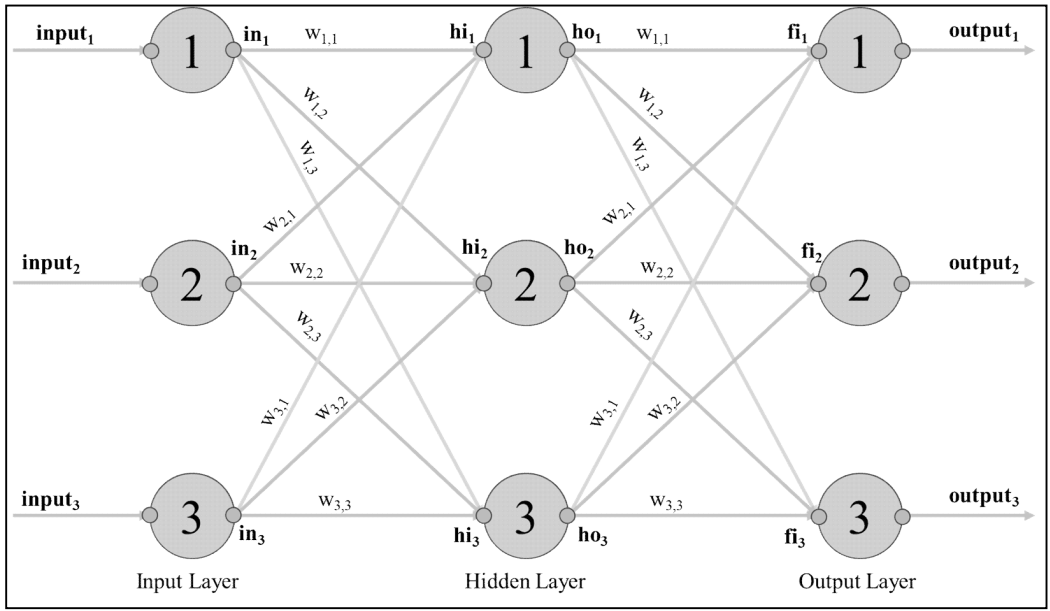

Figure 3: Simple 3x3 Neural Network for image classification problem.

IV. PYTHON SIMULATION

After making the three optimizations described and calculating the accuracy of the new neural network, we expanded all the vectorized computations into defined arithmetic operations.

Just to reiterate, the mechanism we are employing for implementing a neural network on an FPGA is to auto generate the neural network Verilog file from our python script. The reason for doing this is so that any trained neural network can be loaded onto an FPGA. To generate the file, instead of immediately creating a Verilog file, we generated a pseudo-Verilog file that is written in python, such that every line of code can be translated directly into Verilog. The reason for this is so that we can test to make sure the method we are using is correct and still produces a 92% accuracy.

```
# scale input to either 0 or 1
(1)  in₁ = step(input₁)
(2)  in₂ = step(input₂)
(3)  in₃ = step(input₃)

# derive hidden input value
(4)  hi₁ = w₁,₁*in₁ + w₂,₁*in₂ + w₃,₁*in₃
(5)  hi₂ = w₁,₂*in₁ + w₂,₂*in₂ + w₃,₂*in₃
(6)  hi₃ = w₁,₃*in₁ + w₂,₃*in₂ + w₃,₃*in₃

# scale hidden output value to 0 or 1
(7)  ho₁ = step(hi₁)
(8)  ho₂ = step(hi₂)
(9)  ho₃ = step(hi₃)

# derive final input value
(10) fi₁ = w₁,₁*ho₁ + w₂,₁*ho₂ + w₃,₁*ho₃
(11) fi₂ = w₁,₂*ho₁ + w₂,₂*ho₂ + w₃,₂*ho₃
(12) fi₃ = w₁,₃*ho₁ + w₂,₃*ho₂ + w₃,₃*ho₃

# predict final output
(13) prediction = maximum (fi₁, fi₂, fi₃)
```

Figure 4: Python Script 3x3 NN Code Segment

In this autogenerated Verilog file, all vector multiplication is expanded so that every intermediate step for transforming the input into a prediction is explicit. Figure four above shows a simple 3x3 neural network.

In this figure, there are easily identifiable nodes. First are the three input nodes. These nodes are on the far left of the input layer labeled 1, 2, and 3. These inputs are representative of an images color value in the range of 0 – 256. The node on the opposite side of the input node labeled "in" is the adjusted input value after being transformed to either 0 or 1 where the threshold input value is 128.

The next layer has both hidden input (hi) and hidden output (ho) nodes. The hidden input nodes are calculated by summing the weighted inputs. The hidden outputs are calculated by using the activation function on the hidden input values. This process repeats for the final layer. The math for computing these values is shown in the Figure 5 code segment.

Expanding each nodal layer is required when implementing the neural network on an FPGA because every variable must be assigned individually. Unlike CPUs, FPGA's can not deal with vectorized addition or multiplication.

A. Testing NN using Reduced Complexity Techniques

The autogenerated python script contains all expanded calculations that take place in this neural network. Upon execution, the result of the autogenerated python script yielded the same results as the neural network using matrix multiplication. The main difference between the expanded neural network and the vectorized neural network is that the execution time is significantly longer because all vectorized calculations were removed. This script can make ~1000 predictions per second.

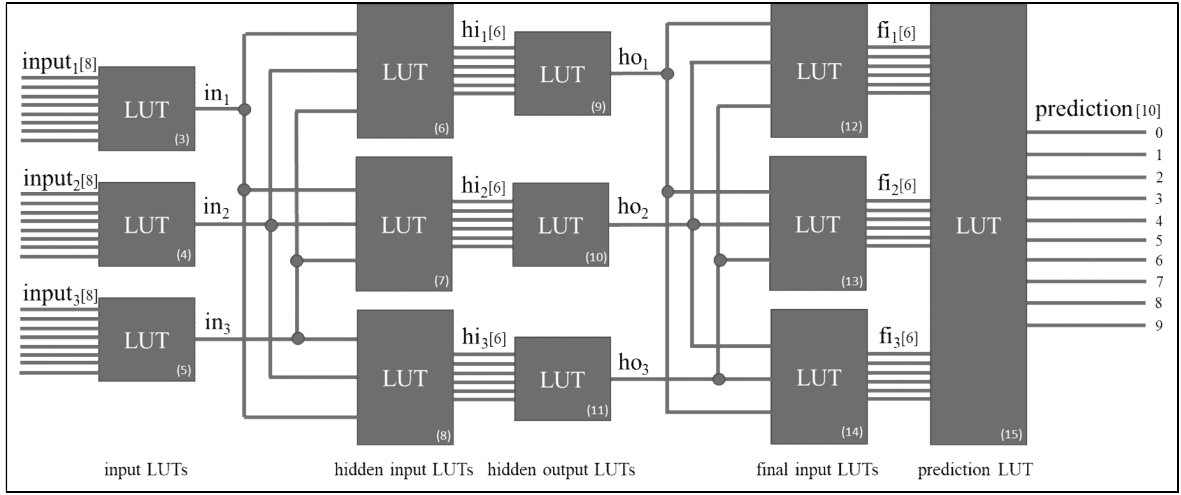

Figure 5: Generalized Hardware Architecture for neural network implentation on an FPGA.

Visually, the code complexity was reduced because the code only contains addition, multiplication, and explicit comparisons, all separated into defined sections.

## V. Implementation on an FPGA

With a simplified neural network where all nodes are fully defined, translating the python code into a hardware defined language is very simple. All variable declaration in python become assign statements using wires. All comparisons such as the step function or final output classification become assigned if statements using the ( ? : ) operators.

### A. Dependent Variable Setup on FPGA

Rewriting the variable assignments in Verilog for the neural network shown in Figure 4, results in the Verilog code in Figure 7.

```
// wire setup
(1)  wire in1, in2, in3, ho1, ho2, ho3;
(2)  wire [5:0] hi1, hi2, hi3, fi1, fi2, fi3, prediction;

// assign scaled inputs
(3)  assign in1 = (input1 > 128) ? 1 : 0;
(4)  assign in2 = (input2 > 128) ? 1 : 0;
(5)  assign in3 = (input3 > 128) ? 1 : 0;

// assign hidden inputs (-10 < weights < 10)
(6)  assign hi1 = w1,1*in1 + w2,1*in2 + w3,1*in3;
(7)  assign hi2 = w1,2*in1 + w2,2*in2 + w3,2*in3;
(8)  assign hi3 = w1,3*in1 + w2,3*in2 + w3,3*in3;

// assign hidden outputs
(9)  assign ho1 = (hi1 > 0) ? 1 : 0;
(10) assign ho2 = (hi2 > 0) ? 1 : 0;
(11) assign ho3 = (hi3 > 0) ? 1 : 0;
// assign final inputs (-10 < weights < 10)
(12) assign fi1 = w1,1*ho1 + w2,1*ho2 + w3,1*ho3;
(13) assign fi2 = w1,2*ho1 + w2,2*ho2 + w3,2*ho3;
(14) assign fi3 = w1,3*ho1 + w2,3*ho2 + w3,3*ho3;

// assign prediction
(15) assign prediction = (fi1 > fi2 && fi1 > fi3) ? 2'b01 :
                        (fi2 > fi3) ?  2'b10 : 2'b11;
```

Figure 6: Verilog 3x3 NN Code Segment

Notice that with this implementation of the 3x3 example on the FPGA, there are no state registers being used. This is important because it means that there are no clocks. Since there is no synchronization clock, the prediction time is only limited by the propagation delay of the logic gates. Furthermore, since all assignments are done asynchronously and in parallel, this is the fastest possible method to form a prediction.

### B. Visualization of Architecture

The structure of the architecture in Figure 6 is defined in the Verilog shown in Figure 7. From this code hardware is produced via a hardware Verilog compiler.

The architecture defined in Figure 7 is interesting because it only contains simple Look Up Tables (LUTs). There is no synchronizing clock or expensive look up table. Every wire shown in the diagram represents a single bit. The largest lookup table used is only 18 bits for classifying the output.

Each simple LUT has an associated line of code in the example above. The line of code is indicated by the number in parenthesis in the bottom right of each LUT symbol. For example, the top left LUT in the "input LUT" column is associated with coding line (3) which sets $in_1$ as either 0 or 1 depending on the $input_1$ being less than or greater than 128. In the "hidden input LUT" column, the lookup tables encode the weight matrix multiplication from lines (6-8).

The "hidden output LUTS" are the hardware representation of coding lines (9-11). The purpose of these LUTs are to determine if the six-bit signed output is positive or negative. If the output is non-negative, the output is 1, otherwise the output is zero.

The "final input LUT" column is the hardware representation of coding lines (12-14). These LUTs are used to encode the second weight matrix. The values returned from these LUTs are six-bit signed outputs.

The "prediction LUT" takes in all three six-bit signed values and calculates which value is the maximum. These comparisons are all encoded in an eighteen input LUT.

This example shows how a simple 3-layer, 3 input neural network is implemented on an FPGA. To accurately represent the neural network used for testing, this architecture must be scaled to implement 784 input LUTs, 500 hidden input LUTs, 500 hidden output LUTs, 10 final input LUTS, and 1 prediction LUT. The results of the actual neural network implementation are explained next.

*C. Initial Results*

Unfortunately, the results for the first run of this code were disappointing. Although the python script shows an accuracy of 92%, the Verilog code that implements this neural network required more logic units than the Xilinx BASYS 3 FPGA is capable of providing.

The FPGA being used to implement this neural network is a Xilinx BASYS 3, Artix 7 FPGA with 33,280 Logic cells. The number of logic cells required for this project was well over 80,000 logic cells. Based on this constraint, further hardware optimizations were required.

*D. Hardware Simplifications*

The root of this problem is that there are too many entries in each lookup table. Despite being very simple LUTs, they still use more logic cells than required.

The first simplification made was for each comparator. Previously, the input LUTs and hidden output LUTs are creating a LUT using every input bit for the comparison table. This is unnecessary. To simplify the input LUT table, only the most significant value must be compared. If the value of the MSB is 1, the output is a 1 and if the MSB is a 0, the output is a 0. This removes the LUT entirely and inputs can be directly passed to the hidden input layer.

Similarly, the hidden output LUTS only need the MSB as well. Since the datatype for the hidden input is a signed wire, if the MSB is a one, then the number is negative so the output of the LUT should be 0. If the MSB is a zero, then the number is positive, so the output of the LUT should be 1. This removes the hidden output LUT entirely and replaces it with an inverter.

Rerunning this code showed a decrease in required logic cells but apparently, the LUTs that were replaced were already highly optimized because the reduction was not significant enough to be able to fit the resulting implementation to our FPGA. Further simplification was required.

Upon further inspection, the LUTs which used the most logic cells were the LUTs encoding the weight matrices. In the simple example from before, line 3 shows the entire calculation for the output of the LUT. This example only has three inputs being scaled by weights and summed. For the actual neural network, there are 784 possible values being scaled by weights and summed. This leads to a massive LUT that needs to be simplified.

The first optimization occurs during Verilog generation from the python script. Taking a closer look at the weight matrices, some of the weight values are zero, thus they do not affect the hidden input and final input values. These weights and inputs can be removed from the calculation. This reduces the number of necessary logic cells by almost 50%, however there are still over 38,000 logic units required.

The second optimization on the hidden and final input LUTs also occurs during Verilog generation. The idea is that multiplication on an FPGA requires significant logic. A simpler implementation is to only use addition. An example is provided below.

```
// w1,1 = 3, w2,1 = 1, w3,1 = 1
(6)   assign hi1 = w1,1*in1 + w2,1*in2 + w3,1*in3;
```

Instead of using multiplication use addition and remove the weights entirely.

```
(6)   assign hi1 = in1 + in1 + in1 + in2 + in3;
```

This simplification reduced the number of required logic cells from over 38,000 to just under 16,000. This number of logic cells is well within the range of the BASYS 3 FPGA. Now that the FPGA is capable of being loaded with the neural network, testing can proceed.

*E. FPGA vs CPU*

The neural network works as expected and is able to classify input images properly with a 92% accuracy using the MNIST dataset. While having the same accuracy as the CPU, the FPGA exhibits significant performance benefits over the CPU.

To begin, the time that it takes for the FPGA to reach a prediction is only limited by the propagation delay of the logic gates. Using this architecture, the propagation delay is ~20 ns.

This is an incredibly fast predictor, however, processing multiple images is limited by the clock on the FPGA. The input data is connected to the neural network via registers. New data to predict must be clocked into the input registers and the fastest that data can be received is limited by the frequency of the FPGA. The maximum frequency of a high-quality FPGA is ~500MHz. This means that this neural network architecture is capable of 500 million predictions per second – pretty incredible.

In comparison, a CPU operating in the GHz frequency range is only capable of a few thousand predictions per second. On an intel core i7 the equivalent python script for the Verilog script can only predict ~1000 predictions per second.

The FPGA implementation of the equivalent neural network is hundreds of thousands of times faster (500 million predictions/sec vs 1000 predictions/sec) than a neural network on a CPU.

VI. CONCLUSION

*A. A More Practical Implementation*

One application for this research is to connect the input for the neural network to a camera module. The camera can output data to the neural network. The prediction for the new image will be ready before the next image is sent to the neural network, as camera image generation is typically never faster than video which is a new image every 33ms.

*B. Scalability*

This architecture is simple yet scalable. Since the Verilog script is generated by a python script, increasing the depth of the neural network is simple. The reason that this neural network is so simple is because all training is done in the python script and the trained neural network is ported to a hardware design to optimize inference performance.

One future advancement for this architecture is to be able to write a python library that captures tensor flow neural network designs and convert them into a hardware defined language architecture.

REFERENCES

[1] Rashid, Tariq, *Make Your Own Neural Network*, USA: CreateSpace Independent Publishing Platform, 2016.
[2] Yufeng, Hao, "A General Neural Network Hardware Architecture on FPGA", University of Birmingham, Edgbaston, Birmingham, B152TE, UK.
[3] M. Nielsen. "Neural Network and Deep Learning". [Online] Available: http:// neuralnetworksanddeeplearning.com/chap3.html
[4] K. Gurney. An Introduction to Neural Networks. 1 edition, CRC Press, pp. 29-30, 1997.
[5] Sahin S., Becerikli Y., Yazici S. (2006) Neural Network Implementation in Hardware Using FPGAs. In: King I., Wang J., Chan LW., Wang D. (eds) Neural Information Processing. ICONIP 2006. Lecture Notes in Computer Science, vol 4234. Springer, Berlin, Heidelberg
[6] Yuteng Zhou and Shrutika Redkar and Xinming Huang, "Deep learning binary neural network on an FPGA," in *2017 IEEE 60th International Midwest Symposium on Circuits and Systems (MWSCAS),* Boston, MA, USA, August 6-9, 2017.